\documentclass[aps,pre,twocolumn,groupedaddress,showpacs]{revtex4-1}

\usepackage[english]{babel}
\usepackage[final]{showkeys} 

\usepackage{dcolumn}   
\usepackage{bm}        

\usepackage{natbib}
\usepackage{graphics}
\usepackage{amsfonts}
\usepackage{amssymb}
\usepackage{amsmath}
\usepackage{textgreek} 

\usepackage[T1]{fontenc}
\usepackage{dcolumn}   
\usepackage{bm}        

\begin{document}



\title{Long-range overcharging and long-range charge reversal in model colloidal dispersions}

\author{Jos\'{e} Alfredo Gonz\'{a}lez-Calder\'{o}n}
\affiliation{CONACyT - Centro de Ingenier\'ia y Desarrollo Industrial (CIDESI), Av. Playa Pie de la Cuesta 702, Desarrollo San Pablo, 76125 Quer\'etaro, Qro., M\'exico.}

\author{Mart\'in Ch\'avez-P\'aez}
\affiliation{Instituto de F\'isica, Universidad Aut\'onoma de San Luis Potos\'i, \'Alvaro Obreg\'on 64,
78000 San Luis Potos\'i, S.L.P., M\'exico.}

\author{Enrique Gonz\'alez-Tovar}
\affiliation{Instituto de F\'isica, Universidad Aut\'onoma de San Luis Potos\'i, \'Alvaro Obreg\'on 64,
78000 San Luis Potos\'i, S.L.P., M\'exico.}

\author{Marcelo Lozada-Cassou}
\email{corresponding author email: marcelolcmx@ier.unam.mx}
\affiliation{Instituto de Energ\'ias Renovables, Universidad Nacional Aut\'onoma de M\'exico (U.N.A.M.), 62580 Temixco,
Mor., Mexico.}

\date{\today}



\begin{abstract}
Most theoretical and simulation studies on charged particles suspensions are at infinite dilution conditions. Hence these studies have been focused on the electrolyte structure around an isolated central particle (or electrode), where phenomena as charge reversal, charge inversion and overcharging have been shown to be relevant.  However, experimental studies at finite volume fraction exhibit interesting phenomenology which imply very long-range correlations. In this paper we apply an integral equation theory to a simple model for a charged macroions suspensions, \textit{at finite volume fraction}, and find two new effects of long-range overcharging and long-range charge reversal. These new effects are different from the classical overcharging and charge reversal in that they occur at finite macroion's volume fraction, far away from the central macroion, are much more intense, and increase, \textit{not decrease}, as a function of the distance to the central particle, which is indicative of correlations at large separations. We find our results to be qualitative consistent with existing experimental results, and Monte Carlo simulations

\end{abstract}


\maketitle


The study of colloidal and nanoparticles interactions and their interfacial properties is a major subject in the fields of physics, chemistry, biology, energy and technology~\cite{wennerstrom,vatamanu}. Studies on charged colloidal suspensions have been made for isolated macroions or two like-charged colloids particles, i.e., at infinite dilution, immersed into a model electrolyte~\cite{verwey48,stillinger1960,schmitz93,Kjellander86,levin99,belloni2000,levin2016}. In theoretical studies of size-symmetrical electrolytes, next to a \textit{single, charged electrode or macroion} (infinite dilution models), oppositely charged ions (counterions) are adsorbed to the surface. Under certain conditions, the charge of these counterions can overcome the surface charge. As a result, the effective electrical field produced by the electrode/colloid plus that due to the counterions reverses its direction with respect to the unscreened electrode electrical field. This phenomenon is known as charge reversal (CR), and has proved to be of impact for electrophoresis experiments~\cite{quesada2002,molina03,quesada03,lobaskin13} and interfacial phenomena~\cite{Wu17}. This effective, reversed electrical field, in turn, produces a layer of electrode's coions, which reverse again the effective electrical field. This second phenomenon is referred to as charge inversion (CI)~\cite{attard96,kjellander98,hsiao08}. Furthermore, if ionic-size asymmetry (or other features, e.g., dielectric contrast) is considered~\cite{kjellander1992,zuckerman2001}, a new phenomenon of \emph{overcharging} (OC)~\cite{jimenez04,ivan10}, also referred to as \emph{surface charge amplification}~\cite{chialvo2008}, occurs, and should not be confused with the same term, used sometimes as a synonymous of charge reversal in the literature~\cite{quesada2002,molina03,lobaskin13,lyklema06}. Overcharging on the surface of an electrode or a central large particle (in a bulk solution) is due to the adsorption of its coions, which are attached to its counterions, brought in contact with the surface of the central particle due to an energy-entropy balance~\cite{jimenez04}. \textit{This short-range OC occurs within a few angstroms next to the electrode's surface}, and has received recent attention in the literature because of its implications on interfacial phenomena\cite{Messina2007,chialvo2008,wang01,chialvo2011,olvera2014,podgornik2014,vatamanu,wang2016,patra2016,seanea2017b,Bhuiyan2017}. While structural oscillations in size-symmetrical ionic solutions were first reported by Kirkwood et al.~\cite{poirier1954,stillinger1960} since 1954, the oscillations of the induced charge have been more recently introduced\cite{deserno2001}, and are presently under active investigation.

Theoretical and simulation investigations of nano-particles suspensions \textit{at finite volume fractions} are technically more difficult due to the large increase of integration space (with different length scales) in theoretical equations, or the to huge number of particles in, necessarily, larger simulation boxes. Thus, many studies of finite concentration colloidal dispersions use some kind of charge renormalization~\cite{belloni1998,bocquet2002,ise99}, consider low added salt concentration, or no added salt~\cite{linse99}. In general these very important investigations are for relatively low volume fraction~\cite{belloni1985,lobaskin03}. A widely used model to study size-asymmetrical electrolytes is the primitive model~\cite{kjellander1992,zuckerman2001,vlachy2011}. 
An extension of this model, to study colloidal dispersions at finite concentration, is the colloidal primitive model (CPM)~\cite{belloni1985,manzanilla2011a,manzanilla2011b,manzanilla2013}, where the particles are taken to be charged, hard spheres, immersed in a continuous solvent of dielectric constant $\epsilon$, and such that the diameter and charge of one of the species is much larger than those of the others. On the other hand, experiments on the structure of polymer latex dispersions exhibit very long-range correlations and coexisting ordered-disordered regions, which imply long-range repulsive and attractive forces~\cite{ise99,tata2008}, which are not yet well understood.

Since the pioneer integral equations of Kirkwood, et al.~\cite{poirier1954,stillinger1960}, for homogeneous and inhomogeneous size-symmetrical electrolytes, based on a density expansion plus a superposition approximation, other approximated integral equations based on the Ornstein-Zernike (OZ) equation~\cite{munster1,McQuarrie} have been derived. Among them is the Hypernetted-Chain/Mean Spherical Approximation (HNC/MSA)~\cite{Lozada-Cassou92a,attard96}. This theory has been extensively, and successfully, compared with density functional theories, MC and/or molecular dynamics (MD) simulations for a large variety of charged homogeneous and inhomogeneous fluids~\cite{vlachy86,vlachy2000,Wu2004,molina2006,olvera2011,manzanilla2011a,vlachy2011,levin2016,seanea2017}. 

In this letter we extend our previous HNC/MSA studies of the CPM~\cite{manzanilla2011a,manzanilla2011b,manzanilla2013} to calculate the charge induced in the fluid, around a central macroion, as a function of the distance to its surface, and show that a long-range overcharging (LROC) and long-range charge reversal (LRCR) can occurs for apropiate conditions. These new phenomena of LROC and LRCR are different from the classical overcharging and charge reversal, briefly discussed above, in that they occur at a finite macroion's volume fraction, far away from the central macroion, are much more intense, and increase, \textit{not decrease}, as the distance to the central particle augments, which imply long-range correlations. We find our results to be qualitatively consistent with existing experimental results, and our Monte Carlo (MC) simulations.


The multi-component HNC equation for a fluid of $n$ species is~\cite{manzanilla2013}

\begin{eqnarray}
\begin{split}
g_{i j}({\bf r}_{21}) = \qquad\qquad \qquad\qquad \qquad\qquad \qquad\qquad \\ \exp \left\{-\beta u_{ij}(\mathbf{r}_{21}) +
  \sum_{l=1}^{n}\rho_{l} \int_{V} h_{i l}({\bf r}_{23})
c_{lj}(\mathbf{r}_{31} )d\mathbf{r}_{3} \right\}, \\
\label{trainee}
\end{split}
\end{eqnarray}

\noindent with $i,j =1,2,\dots,n$, $\rho_l$ is the number density of species $l$, 
$h_{il}(\mathbf{r}_{23}) \equiv g_{il}(\mathbf{r}_{23})-1$ is the total correlation functions for two particles at
 $\mathbf{r}_{2}$ and $\mathbf{r}_{3}$, of species $i$ and $l$, respectively;
with  $\mathbf{r}_{23}= \mathbf{r}_{2}- \mathbf{r}_{3}$. $g_{ij}({\bf r}_{12})$ is the pair correlation function, also referred to as the {\em radial distribution function}, which gives the probability of finding a particle $1$, of species $j$, at the distance $\mathbf{r}_{12}$, from the central particle $2$, of species $i$. To solve Eq.~\eqref{trainee}, a functional dependence of $c_{lj}(\mathbf{r}_{31})$, with $g_{ij}(\mathbf{r}_{12})$ is needed. $c_{lj}(\mathbf{r}_{13})$ is basically an approximation for a quasiparticle, in the context of the many-body theory~\cite{Friedmanbook}, which, here is taken to be given by the MSA, and for which there is an analytical solution~\cite{kazuo77}. Then, Eqs.~\eqref{trainee} become the HNC/MSA equations.

Here we will solve Eqs.~\eqref{trainee} for the CPM, for a three species charged fluid, i.e., positive ions, negative ions and macroions, of species $+, -$, and $M$, respectively. For simplicity, the counterions of the macroions are taken to be equal to the electrolyte anions, since the colloid's charge will be considered to be positive. Species $M$ are large particles so that their diameter is in the colloidal domain. Thus, the ions diameters are $a_{+}=a_{-}=a$. The interaction potential between two particles of species $i$  and $j$, with a separation distance $r$, is given by
 \begin{equation}
 u_{i j}(r)=
 \left \{ \begin{array}{ll}
 \infty &\text{for $r< a_{ij}$}\\
 {\displaystyle\frac{q_{i}q_{j}}{\epsilon r}}& \text{for $r \geq a_{ij}$} 
 \end{array} \right.  \text{with } i,j=+,-,M.
 \label{mar1}
 \end{equation}
$a_{ij} = (a_{i} + a_{j})/2$. The colloid's diameter is $a_M$, and its surface charge density is $\sigma_0\equiv\sigma_M$$=z_M e/(\pi a_M^2)$. The charge on the colloidal particles' surface, $Q_0\equiv Q_M=\pi a_M^2\sigma_M$, is compensated by the induced charge in the fluid. Hence, as $r\rightarrow\infty$, $Q(r) \rightarrow 0$, where $Q(r)=- 4\pi \int_{r}^{\infty}\rho_{el}(t)t^{2}dt$, and $\rho_{el}(r) \equiv \sum_{j=+,-,M} ez_{j}\rho_{j}g_{M j}(r)$ is the charge density profile.  A detailed derivation of Eqs.~\eqref{trainee}, as well as the electrostatics of the CPM, can be found in reference \cite{manzanilla2013}.



 

To test our HNC/MSA results we performed MC simulations for the CPM. They were conducted in a cubic simulation
box with periodic boundary conditions. The number of particles of the different components was selected considering the size of the box, the volume fraction of the macroions, the salt concentration, and the electroneutrality condition ($\sum_{j=+,-,M} ez_{j}\rho_{j}=0$). The energy of the systems was calculated using the Ewald summation method with a cutoff $R=L/2$ and screening parameter $\alpha=6/L$, where $L$ is the length of the unit cell~\cite{lobaskin99}. The systems were evolved using the standard displacement of individual ions, complemented by cluster moves, a refined technique~\cite{lobaskin99} where full clusters of particles (i.e. a macroion plus some of the small ions) are displaced in a single move, facilitating the sampling of the configurational space in systems of particles that tend to form aggregates; our implementation of cluster moves follows closely the details presented in Ref.~\cite{lobaskin99}.


With the radial distribution functions obtained from Eqs.~\eqref{trainee} we calculate the normalized induced charge, as a function of the distance to the surface of the central macroion, $Q(r)/Q_0$, for several colloidal volume fractions, $ \phi\equiv\frac{1}{6}\pi\rho_Ma_M^3$, and surface charge densities $\sigma_0$. In all cases the added salt is a 1:1, $0.1\,M$ electrolyte, with ionic diameter, $a=4.25\,\mbox{\AA}$. The macroions diameter is $a_M=10a$, and $\epsilon=78.5$. Both the electrolyte ions and macroions are assumed to have the same dielectric constant to avoid image charges. The temperature $T=298ºK$. 

\begin{figure}
\resizebox{0.5\textwidth}{!}{\includegraphics{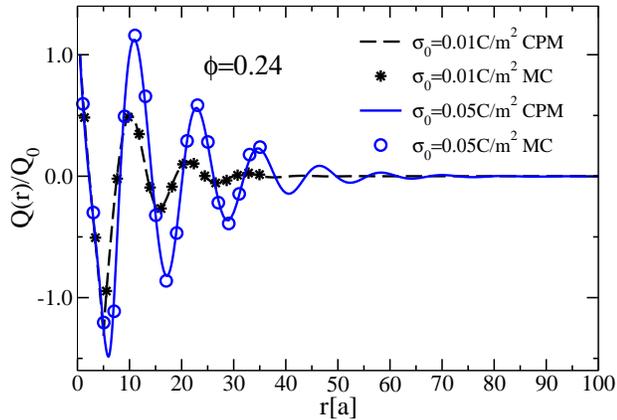}}
\caption{\label{Fig1} HNC/MSA and MC results for ${Q(r)/Q_0}$, for the CPM, as a function of the distance from the central colloid particle surface, and for two different colloidal surface charge densities. The symbols are the MC simulations and the lines are the HNC/MSA results.}
\end{figure}

In Fig.~\ref{Fig1} we compare the HNC/MSA and MC normalized induced charge, $Q(r)/Q_0$. The agreement is very good. But, more important, notice that, for $\sigma_0=0.05\,C/m^2$, $Q(r)/Q_0\geqslant1$, for $r\approx 11a$, away from the central macroparticle's surface. This implies that there is an effective charge, above the value of the macroparticle's original charge, at some distance away from its surface. This new LROC differs from the previously reported OC~\cite{jimenez04}, in that the LROC occurs far from the central particle, and the usual OC only next to the central macroparticle.  In addition to the LROC, we observe an important \textit{charge reversal} (of around 1.5 the original charge), at $r\approx 7a$, of a magnitude higher, in absolute value, than the original charge. While this phenomenon has been reported before for a CPM next to a charged plate~\cite{jimenez04}, and very recently observed for size-symmetic electrolytes, also next to a charged plate, when considering the solvent and the plate with different dielectric constants~\cite{Wu17}, the strong charge reversal reported here has not been published before to occur far from the central particle, as it is seen in Fig.~\ref{Fig1}, for $r\approx 17a$. For lower volume fractions and/or particles charge, the induced charge is less intense, as can be seen in Fig.~\ref{Fig1}, for the $\sigma_0=0.01\,C/m^2$ case. However the long-range correlation will be present up to relatively low values of volume fraction. In the present calculations, we went as low as 0.06 of volume fraction, and for sufficiently low macroions charge the charge oscillations are very mild. They disappear, of course, for cero macroions valence.

\begin{figure}
\resizebox{0.5\textwidth}{!}{\includegraphics{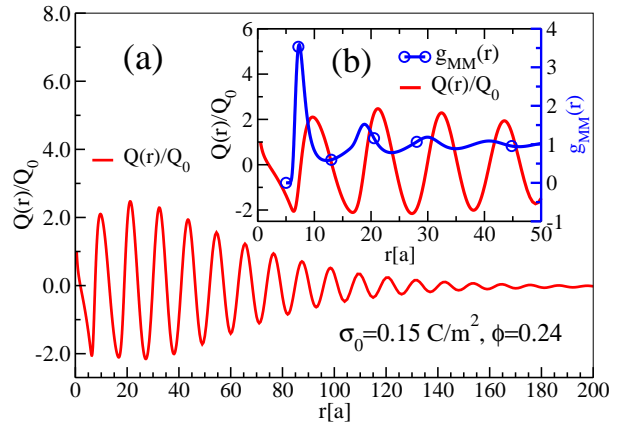}}
\caption{\label{Fig2} (a) HNC/MSA results for $Q(r)/Q_0$, as a function of the distance from the central colloid particle surface, for a high value of $\sigma_0$. (b) Comparison of the macroion-macroion radial distribution function, $g_{MM}(r)$, with its corresponding, reduced induced charge $Q(r)/Q_0$ profile.}
\end{figure}

\begin{figure}
\resizebox{0.47\textwidth}{!}{\includegraphics{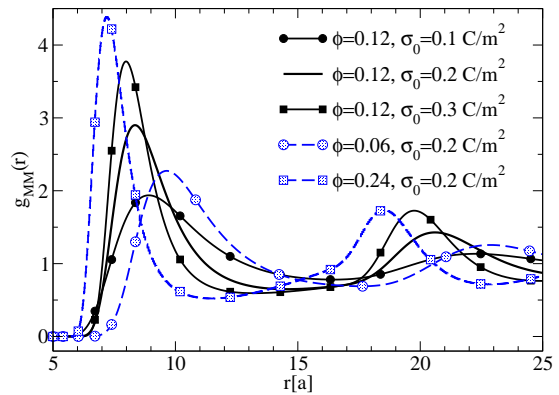}}
\caption{\label{Fig3}  HNC/MSA radial distribution functions for macroions, $g_{MM}(r)$, as a function of the distance to the central macroion surface. Two sets of curves are displayed: One set is for a constant volume fraction $\phi=0.12$ and three values of the macroions surface charge density ($\sigma_0=0.1, 0.2$ and $0.3\, C/m^2$), and another set for a constant macroion surface charge density $\sigma_0=0.2\, C/m^2$ and three values of the macroions volume fraction $\phi= 0.06, \,0.12$ and $0.24$. The curve for $\phi=0.12$ and $\sigma_0=0.2\, C/m^2$ is common to the two sets of curves.}
\end{figure}

In Fig.~\ref{Fig2}, we portray $Q(r)/Q_0$, for $\sigma_0=0.15$ $C/m^2$. At this higher surface charge density, the induced LROC is more than twice the original charge of the central macroion, i.e., $Q(r)/Q_0\geqslant2$, between $9a\leqslant r \leqslant33a$, and $Q(r)/Q_0\geqslant 1$ as far as at ${\sim}65a$, measured from the central macroion surface. The charge correlation, however, extends up to around $300a$. For $\sigma_0=0.3$ $C/m^2$ and $\phi=0.24$ (not shown), the LROC goes as far as $r=190a$, and the \textit{charge correlation to around $800a$}. In the standard interfacial science argot the first minimum would correspond to a CR and the first maximum to a CI, and both have an absolute value lower than the charge on the the central particle or electrode. However, in Fig.~\ref{Fig2} we see a very different phenomenology, i.e., the first minimum more than doubles the original charge $Q_0$, i.e., $Q(r)/Q_0\approx-2.1$, for $r\approx6.4a$ (perhaps a more giant charge inversion than that reported before~\cite{jimenez04,Wu17}), and the first maximum also is more than twice the $Q_0$, i.e., $Q(r)/Q_0\approx +2.1$. Moreover, \textit{the second maximum is even larger the first maximum, and the second minimum is lower than the first minimum. For  $\sigma_0=0.3$ $C/m^2$ the third maximum and minimum are even larger than their corresponding second maximum and minimum}. Hence, for homogeneous finite concentration colloidal dispersions the concepts of CR and CI are probably not appropriate, and instead hereinafter we will refer to the charge reversals reported here, present at every distance from the central particle, as long-range charge reversal (LRCR), and we will not use the concept of charge inversion. In the inset of Fig.~\ref{Fig2}, we have superimposed the macroion-macroion radial distribution function, $g_{MM}(r)$ to the $Q(r)/Q_0$ curve. As expected, the maxima and minima in $g_{MM}(r)$, which indicates the most probable position of macroions around the central particle, are closely correlated with the maxima and minima of $Q(r)/Q_0$, i.e., with the $r$-interval where the LROC and LRCR occur. We have calculated $Q(r)/Q_0$ for an interval of $0.06\leqslant\phi\leqslant 0.24$ and $0\leqslant \sigma_0 \leqslant0.3$ $C/m^2$, and, in general, the LROC and LRCR disappears for a combination of low values of $\phi$ and $\sigma_0$. Our results on the LROC and/or LRCR for charged macroions, suggest that under the action of a external field, the macroions might move as a cluster of particles. If a lower dielectric constant of the macroparticles, than that of the solvent is considered, the range of the electrical field in the macroions solution will increase, and, hence, probably also the intensity and range of the long-range charge correlation. 

In Fig.~\ref{Fig3}, we show the $g_{MM}(r)$, for two sequences of results: in one we keep the central particle surface charge density constant, at $\sigma_{0}=0.2$ $C/m^2$, and vary the volume fraction. In the other, we keep $\phi=0.12$ constant, and vary the macroparticle's surface charge density. For the case in which we keep the charge density constant, we see that, as $\phi$ increases, the position of the maxima in $g_{MM}(r)$ becomes closer to the central particle, as expected, since a higher volume fraction reduces the available volume, and hence, apparently, the attractive entropic force overcomes the repulsive coulombic force. We further discuss this point below. On the other hand, if we keep the volume fraction constant $\phi=0.12$, counterintuitively we see that as the colloid's charge increases, the macroions become closer to each other, i.e., in spite that a higher charge implies a higher coulombic force repulsion, and that apparently the entropic force has been kept constant (since $\phi$ is constant).  The explanation resides in the structure of Eq.~\ref{trainee}, which is a nonlinear equation that entangles both coulombic and entropic forces. Thus, a higher charge, implies a higher repulsion, which, in turn, produces a lower available volume. Therefore, the gathering of the macroions, with increasing macroions charge, is an overall entropy effect. We observed this same behavior for $\phi=0.06$ and $0.24$, and even for relatively low values of the macroions charge, e.g., for $\sigma_{0} \approx 0.05 C/m^2$ and higher.  A qualitatively  stronger effective long-range attraction for higher charge density particles has been experimentally observed by Ise, et al.~\cite{ise83,ise99}, in structure studies of polymer latex particles. Our results seem to qualitatively agree with these experimental results. 

\begin{figure}
\resizebox{0.47\textwidth}{!}{\includegraphics{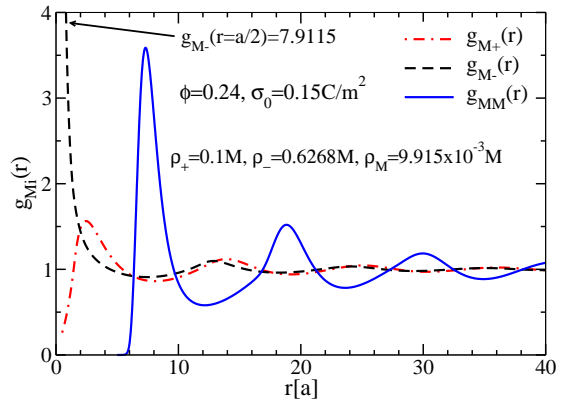}}
\caption{\label{Fig4} Radial distribution functions for anions, $g_{M-}(r)$, cations, $g_{M+}(r)$, and macroions, $g_{MM}(r)$, as a function of the distance to the central macroion surface. The corresponding bulk concentrations,  $\rho_+$, $\rho_-$ and $\rho_M$ are displayed for easy reference.}
\end{figure}

To complete the picture of the origin of the LROC and LRCR, in Fig.~\ref{Fig4}, we show the ions and macroions radial distribution functions. We point out that in-between every two successive maxima of $g_{MM}$, the anions and cations radial distribution functions are \textit{above} of their bulk values and practically overlap. Notice that, because of the presence of the macroions' counterions, $\rho_{-}(r)\equiv\rho_{-}^{bulk}g_{-}(r)$ is much larger than $\rho_{+}(r)\equiv\rho_{+}^{bulk}g_{+}(r)$. In Fig.~\ref{Fig4}, $\rho_{+}^{bulk}\equiv \rho_{+}=0.1\,M$, and $\rho_{-}^{bulk}\equiv \rho_{-}=0.6286\,M$. Thus, in some degree, the macroions, with their surrounding electrical double layers, behave as hard spheres, with the consequent decrease in the available volume, and the increase of the entropy force. Another possible explanation of the unexpected colloidal attraction, suggested in the past~\cite{ise99}, proposes that the anions in between the like-charged macroions, electrostatically attract them, overcoming their original repulsion, and are responsible for a long-range attractive force, necessary to explain the observed order-disorder phase coexistence in polymer latex experiments. To elucidate this point, we have calculated the total, electrical and entropic components of the macroion-macroion force (not shown), and we found that, while both the entropic and electrical forces are oscillatory, and of very long-range, it is the entropic force which always dominates to produce both the net attraction and repulsion between the macroions, at every distance.
 
In summary, two new phenomena are predicted, for bulk macroions dispersions, at finite concentration: a long-range overcharging (LROC) and a long-range charge reversal (LRCR). These two phenomena are different from the classical overcharging and charge reversal in that they occur at finite macroion's volume fraction, far away from the central macroion, are much more intense, and increase, \textit{not decrease}, as a function of the distance to the central particle, which is a signal of correlations at large separations. In particular, they seem to be consistent with the long-range correlation experimentally observed in polymer latex dispersions, where the aggregation of like-charged particles, as the charge increases, is observed, and, as in the experiments, our results imply long-range correlation among the macroions. In particular, it is shown that at finite volume fraction the colloidal dispersion is ruled by a long-range energy-entropy balance. However, the long-range attractions and repulsions are entropy effects, which overcome the electrostatic repulsion of the like-charged colloid particles. The long-range nature of the induced charge around the central particle, could have relevant implications for electrophoresis experiments.
\noindent\textbf{Acknowledgments} The support of CONACYT (Grant No. 169125) is gratefully acknowledged.


\begin{thebibliography}{58}%
\makeatletter
\providecommand \@ifxundefined [1]{%
 \@ifx{#1\undefined}
}%
\providecommand \@ifnum [1]{%
 \ifnum #1\expandafter \@firstoftwo
 \else \expandafter \@secondoftwo
 \fi
}%
\providecommand \@ifx [1]{%
 \ifx #1\expandafter \@firstoftwo
 \else \expandafter \@secondoftwo
 \fi
}%
\providecommand \natexlab [1]{#1}%
\providecommand \enquote  [1]{``#1''}%
\providecommand \bibnamefont  [1]{#1}%
\providecommand \bibfnamefont [1]{#1}%
\providecommand \citenamefont [1]{#1}%
\providecommand \href@noop [0]{\@secondoftwo}%
\providecommand \href [0]{\begingroup \@sanitize@url \@href}%
\providecommand \@href[1]{\@@startlink{#1}\@@href}%
\providecommand \@@href[1]{\endgroup#1\@@endlink}%
\providecommand \@sanitize@url [0]{\catcode `\\12\catcode `\$12\catcode
  `\&12\catcode `\#12\catcode `\^12\catcode `\_12\catcode `\%12\relax}%
\providecommand \@@startlink[1]{}%
\providecommand \@@endlink[0]{}%
\providecommand \url  [0]{\begingroup\@sanitize@url \@url }%
\providecommand \@url [1]{\endgroup\@href {#1}{\urlprefix }}%
\providecommand \urlprefix  [0]{URL }%
\providecommand \Eprint [0]{\href }%
\providecommand \doibase [0]{http://dx.doi.org/}%
\providecommand \selectlanguage [0]{\@gobble}%
\providecommand \bibinfo  [0]{\@secondoftwo}%
\providecommand \bibfield  [0]{\@secondoftwo}%
\providecommand \translation [1]{[#1]}%
\providecommand \BibitemOpen [0]{}%
\providecommand \bibitemStop [0]{}%
\providecommand \bibitemNoStop [0]{.\EOS\space}%
\providecommand \EOS [0]{\spacefactor3000\relax}%
\providecommand \BibitemShut  [1]{\csname bibitem#1\endcsname}%
\let\auto@bib@innerbib\@empty
\bibitem [{\citenamefont {Evans}\ and\ \citenamefont
  {Wennerstr{\"o}m}(1999)}]{wennerstrom}%
  \BibitemOpen
  \bibfield  {author} {\bibinfo {author} {\bibfnamefont {D.~F.}\ \bibnamefont
  {Evans}}\ and\ \bibinfo {author} {\bibfnamefont {H.}~\bibnamefont
  {Wennerstr{\"o}m}},\ }\href@noop {} {\emph {\bibinfo {title} {The Colloidal
  Domain: Where Physics, Chemistry, Biology, and Technology Meet}}},\ \bibinfo
  {edition} {2nd}\ ed.\ (\bibinfo  {publisher} {Wiley-VCH},\ \bibinfo {address}
  {New York},\ \bibinfo {year} {1999})\BibitemShut {NoStop}%
\bibitem [{\citenamefont {Vatamanu}\ and\ \citenamefont
  {Bedrov}(2015)}]{vatamanu}%
  \BibitemOpen
  \bibfield  {author} {\bibinfo {author} {\bibfnamefont {J.}~\bibnamefont
  {Vatamanu}}\ and\ \bibinfo {author} {\bibfnamefont {D.}~\bibnamefont
  {Bedrov}},\ }\href@noop {} {\bibfield  {journal} {\bibinfo  {journal} {J.
  Phys. Chem. Lett.}\ }\textbf {\bibinfo {volume} {6}},\ \bibinfo {pages}
  {3594} (\bibinfo {year} {2015})}\BibitemShut {NoStop}%
\bibitem [{\citenamefont {Verwey}\ and\ \citenamefont
  {Overbeek}(1948)}]{verwey48}%
  \BibitemOpen
  \bibfield  {author} {\bibinfo {author} {\bibfnamefont {E.~J.~W.}\
  \bibnamefont {Verwey}}\ and\ \bibinfo {author} {\bibfnamefont {J.~T.~G.}\
  \bibnamefont {Overbeek}},\ }\href@noop {} {\emph {\bibinfo {title} {Theory of
  the Stability of Lyophobic Colloids}}}\ (\bibinfo  {publisher} {Dover},\
  \bibinfo {address} {New York},\ \bibinfo {year} {1948})\BibitemShut {NoStop}%
\bibitem [{\citenamefont {Stillinger}\ and\ \citenamefont
  {Kirkwood}(1960)}]{stillinger1960}%
  \BibitemOpen
  \bibfield  {author} {\bibinfo {author} {\bibfnamefont {F.~H.}\ \bibnamefont
  {Stillinger}}\ and\ \bibinfo {author} {\bibfnamefont {J.~G.}\ \bibnamefont
  {Kirkwood}},\ }\href@noop {} {\bibfield  {journal} {\bibinfo  {journal} {J.
  Chem. Phys.}\ }\textbf {\bibinfo {volume} {33}},\ \bibinfo {pages} {1282}
  (\bibinfo {year} {1960})}\BibitemShut {NoStop}%
\bibitem [{\citenamefont {Schmitz}(1993)}]{schmitz93}%
  \BibitemOpen
  \bibfield  {author} {\bibinfo {author} {\bibfnamefont {K.~S.}\ \bibnamefont
  {Schmitz}},\ }\href@noop {} {\emph {\bibinfo {title} {Macroions in Solution
  and Colloidal Suspensions}}}\ (\bibinfo  {publisher} {VCH Publishers},\
  \bibinfo {address} {New York},\ \bibinfo {year} {1993})\BibitemShut {NoStop}%
\bibitem [{\citenamefont {Kjellander}\ and\ \citenamefont
  {Mar\v{c}elja}(1986)}]{Kjellander86}%
  \BibitemOpen
  \bibfield  {author} {\bibinfo {author} {\bibfnamefont {R.}~\bibnamefont
  {Kjellander}}\ and\ \bibinfo {author} {\bibfnamefont {S.}~\bibnamefont
  {Mar\v{c}elja}},\ }\href@noop {} {\bibfield  {journal} {\bibinfo  {journal}
  {Chem. Phys. Lett.}\ }\textbf {\bibinfo {volume} {127}},\ \bibinfo {pages}
  {402} (\bibinfo {year} {1986})}\BibitemShut {NoStop}%
\bibitem [{\citenamefont {Diehl}\ \emph {et~al.}(1999)\citenamefont {Diehl},
  \citenamefont {Tamashiro}, \citenamefont {Barbosa},\ and\ \citenamefont
  {Levin}}]{levin99}%
  \BibitemOpen
  \bibfield  {author} {\bibinfo {author} {\bibfnamefont {A.}~\bibnamefont
  {Diehl}}, \bibinfo {author} {\bibfnamefont {M.~N.}\ \bibnamefont
  {Tamashiro}}, \bibinfo {author} {\bibfnamefont {M.~C.}\ \bibnamefont
  {Barbosa}}, \ and\ \bibinfo {author} {\bibfnamefont {Y.}~\bibnamefont
  {Levin}},\ }\href@noop {} {\bibfield  {journal} {\bibinfo  {journal} {Physica
  A}\ }\textbf {\bibinfo {volume} {274}},\ \bibinfo {pages} {433} (\bibinfo
  {year} {1999})}\BibitemShut {NoStop}%
\bibitem [{\citenamefont {Belloni}(2000)}]{belloni2000}%
  \BibitemOpen
  \bibfield  {author} {\bibinfo {author} {\bibfnamefont {L.}~\bibnamefont
  {Belloni}},\ }\href@noop {} {\bibfield  {journal} {\bibinfo  {journal} {J.
  Phys.: Condens. Matter}\ }\textbf {\bibinfo {volume} {12}},\ \bibinfo {pages}
  {R549} (\bibinfo {year} {2000})}\BibitemShut {NoStop}%
\bibitem [{\citenamefont {Colla}\ \emph {et~al.}(2016)\citenamefont {Colla},
  \citenamefont {Girotto}, \citenamefont {dos Santos},\ and\ \citenamefont
  {Levin}}]{levin2016}%
  \BibitemOpen
  \bibfield  {author} {\bibinfo {author} {\bibfnamefont {T.}~\bibnamefont
  {Colla}}, \bibinfo {author} {\bibfnamefont {M.}~\bibnamefont {Girotto}},
  \bibinfo {author} {\bibfnamefont {A.~P.}\ \bibnamefont {dos Santos}}, \ and\
  \bibinfo {author} {\bibfnamefont {Y.}~\bibnamefont {Levin}},\ }\href@noop {}
  {\bibfield  {journal} {\bibinfo  {journal} {J. Chem. Phys.}\ }\textbf
  {\bibinfo {volume} {145}},\ \bibinfo {pages} {094704} (\bibinfo {year}
  {2016})}\BibitemShut {NoStop}%
\bibitem [{\citenamefont {Quesada-P{\'e}rez}\ \emph {et~al.}(2002)\citenamefont
  {Quesada-P{\'e}rez}, \citenamefont {Mart{\'i}n-Molina}, \citenamefont
  {Galisteo-Gonz{\'a}lez},\ and\ \citenamefont
  {Hidalgo-{\'A}lvarez}}]{quesada2002}%
  \BibitemOpen
  \bibfield  {author} {\bibinfo {author} {\bibfnamefont {M.}~\bibnamefont
  {Quesada-P{\'e}rez}}, \bibinfo {author} {\bibfnamefont {A.}~\bibnamefont
  {Mart{\'i}n-Molina}}, \bibinfo {author} {\bibfnamefont {F.}~\bibnamefont
  {Galisteo-Gonz{\'a}lez}}, \ and\ \bibinfo {author} {\bibfnamefont
  {R.}~\bibnamefont {Hidalgo-{\'A}lvarez}},\ }\href@noop {} {\bibfield
  {journal} {\bibinfo  {journal} {Mol. Phys.}\ }\textbf {\bibinfo {volume}
  {100}},\ \bibinfo {pages} {3029} (\bibinfo {year} {2002})}\BibitemShut
  {NoStop}%
\bibitem [{\citenamefont {Molina}\ \emph {et~al.}(2003)\citenamefont {Molina},
  \citenamefont {Quesada-P{\'e}rez}, \citenamefont {Galisteo-Gonz{\'a}lez},\
  and\ \citenamefont {Hidalgo-{\'A}lvarez}}]{molina03}%
  \BibitemOpen
  \bibfield  {author} {\bibinfo {author} {\bibfnamefont {A.~M.}\ \bibnamefont
  {Molina}}, \bibinfo {author} {\bibfnamefont {M.}~\bibnamefont
  {Quesada-P{\'e}rez}}, \bibinfo {author} {\bibfnamefont {F.}~\bibnamefont
  {Galisteo-Gonz{\'a}lez}}, \ and\ \bibinfo {author} {\bibfnamefont
  {R.}~\bibnamefont {Hidalgo-{\'A}lvarez}},\ }\href@noop {} {\bibfield
  {journal} {\bibinfo  {journal} {J. Chem. Phys.}\ }\textbf {\bibinfo {volume}
  {118}},\ \bibinfo {pages} {4183} (\bibinfo {year} {2003})}\BibitemShut
  {NoStop}%
\bibitem [{\citenamefont {Quesada-P{\'e}rez}\ \emph {et~al.}(2003)\citenamefont
  {Quesada-P{\'e}rez}, \citenamefont {Gonz{\'a}lez-Tovar}, \citenamefont
  {Mart{\'i}n-Molina}, \citenamefont {Lozada-Cassou},\ and\ \citenamefont
  {Hidalgo-{\'A}lvarez}}]{quesada03}%
  \BibitemOpen
  \bibfield  {author} {\bibinfo {author} {\bibfnamefont {M.}~\bibnamefont
  {Quesada-P{\'e}rez}}, \bibinfo {author} {\bibfnamefont {E.}~\bibnamefont
  {Gonz{\'a}lez-Tovar}}, \bibinfo {author} {\bibfnamefont {A.}~\bibnamefont
  {Mart{\'i}n-Molina}}, \bibinfo {author} {\bibfnamefont {M.}~\bibnamefont
  {Lozada-Cassou}}, \ and\ \bibinfo {author} {\bibfnamefont {R.}~\bibnamefont
  {Hidalgo-{\'A}lvarez}},\ }\href@noop {} {\bibfield  {journal} {\bibinfo
  {journal} {Chem. Phys. Chem}\ }\textbf {\bibinfo {volume} {4}},\ \bibinfo
  {pages} {234} (\bibinfo {year} {2003})}\BibitemShut {NoStop}%
\bibitem [{\citenamefont {Semenov}\ \emph {et~al.}(2013)\citenamefont
  {Semenov}, \citenamefont {Raafatnia}, \citenamefont {Sega}, \citenamefont
  {Lobaskin}, \citenamefont {Holm},\ and\ \citenamefont {Kremer}}]{lobaskin13}%
  \BibitemOpen
  \bibfield  {author} {\bibinfo {author} {\bibfnamefont {I.}~\bibnamefont
  {Semenov}}, \bibinfo {author} {\bibfnamefont {S.}~\bibnamefont {Raafatnia}},
  \bibinfo {author} {\bibfnamefont {M.}~\bibnamefont {Sega}}, \bibinfo {author}
  {\bibfnamefont {V.}~\bibnamefont {Lobaskin}}, \bibinfo {author}
  {\bibfnamefont {C.}~\bibnamefont {Holm}}, \ and\ \bibinfo {author}
  {\bibfnamefont {F.}~\bibnamefont {Kremer}},\ }\href@noop {} {\bibfield
  {journal} {\bibinfo  {journal} {Phys. Rev. E.}\ }\textbf {\bibinfo {volume}
  {87}},\ \bibinfo {pages} {022302} (\bibinfo {year} {2013})}\BibitemShut
  {NoStop}%
\bibitem [{\citenamefont {Wang}\ and\ \citenamefont {Wu}(2017)}]{Wu17}%
  \BibitemOpen
  \bibfield  {author} {\bibinfo {author} {\bibfnamefont {Z.-Y.}\ \bibnamefont
  {Wang}}\ and\ \bibinfo {author} {\bibfnamefont {J.}~\bibnamefont {Wu}},\
  }\href@noop {} {\bibfield  {journal} {\bibinfo  {journal} {J. Chem. Phys.}\
  }\textbf {\bibinfo {volume} {147}},\ \bibinfo {pages} {024703} (\bibinfo
  {year} {2017})}\BibitemShut {NoStop}%
\bibitem [{\citenamefont {Attard}(1996)}]{attard96}%
  \BibitemOpen
  \bibfield  {author} {\bibinfo {author} {\bibfnamefont {P.}~\bibnamefont
  {Attard}},\ }in\ \href@noop {} {\emph {\bibinfo {booktitle} {Advances in
  Chemical Physics}}},\ Vol.\ \bibinfo {volume} {XCII},\ \bibinfo {editor}
  {edited by\ \bibinfo {editor} {\bibfnamefont {I.}~\bibnamefont {Prigogine}}\
  and\ \bibinfo {editor} {\bibfnamefont {S.~A.}\ \bibnamefont {Rice}}}\
  (\bibinfo  {publisher} {John Wiley and Sons, Inc.},\ \bibinfo {address} {New
  York},\ \bibinfo {year} {1996})\BibitemShut {NoStop}%
\bibitem [{\citenamefont {Greberg}\ and\ \citenamefont
  {Kjellander}(1998)}]{kjellander98}%
  \BibitemOpen
  \bibfield  {author} {\bibinfo {author} {\bibfnamefont {H.}~\bibnamefont
  {Greberg}}\ and\ \bibinfo {author} {\bibfnamefont {R.}~\bibnamefont
  {Kjellander}},\ }\href@noop {} {\bibfield  {journal} {\bibinfo  {journal} {J.
  Chem. Phys.}\ }\textbf {\bibinfo {volume} {108}},\ \bibinfo {pages} {2940}
  (\bibinfo {year} {1998})}\BibitemShut {NoStop}%
\bibitem [{\citenamefont {Hsiao}(2008)}]{hsiao08}%
  \BibitemOpen
  \bibfield  {author} {\bibinfo {author} {\bibfnamefont {P.-Y.}\ \bibnamefont
  {Hsiao}},\ }\href@noop {} {\bibfield  {journal} {\bibinfo  {journal} {J.
  Phys. Chem. B}\ }\textbf {\bibinfo {volume} {112}},\ \bibinfo {pages} {7347}
  (\bibinfo {year} {2008})}\BibitemShut {NoStop}%
\bibitem [{\citenamefont {Greberg}\ and\ \citenamefont
  {Kjellander}(1992)}]{kjellander1992}%
  \BibitemOpen
  \bibfield  {author} {\bibinfo {author} {\bibfnamefont {H.}~\bibnamefont
  {Greberg}}\ and\ \bibinfo {author} {\bibfnamefont {R.}~\bibnamefont
  {Kjellander}},\ }\href@noop {} {\bibfield  {journal} {\bibinfo  {journal}
  {Chem. Phys. Letts.}\ }\textbf {\bibinfo {volume} {200}},\ \bibinfo {pages}
  {76} (\bibinfo {year} {1992})}\BibitemShut {NoStop}%
\bibitem [{\citenamefont {Zuckerman}\ \emph {et~al.}(2001)\citenamefont
  {Zuckerman}, \citenamefont {Fisher},\ and\ \citenamefont
  {Bekiranov}}]{zuckerman2001}%
  \BibitemOpen
  \bibfield  {author} {\bibinfo {author} {\bibfnamefont {D.~M.}\ \bibnamefont
  {Zuckerman}}, \bibinfo {author} {\bibfnamefont {M.~E.}\ \bibnamefont
  {Fisher}}, \ and\ \bibinfo {author} {\bibfnamefont {S.}~\bibnamefont
  {Bekiranov}},\ }\href@noop {} {\bibfield  {journal} {\bibinfo  {journal}
  {Phys. Rev. E}\ }\textbf {\bibinfo {volume} {64}},\ \bibinfo {pages} {011206}
  (\bibinfo {year} {2001})}\BibitemShut {NoStop}%
\bibitem [{\citenamefont {Jim{\'e}nez-{\'A}ngeles}\ and\ \citenamefont
  {Lozada-Cassou}(2004)}]{jimenez04}%
  \BibitemOpen
  \bibfield  {author} {\bibinfo {author} {\bibfnamefont {F.}~\bibnamefont
  {Jim{\'e}nez-{\'A}ngeles}}\ and\ \bibinfo {author} {\bibfnamefont
  {M.}~\bibnamefont {Lozada-Cassou}},\ }\href@noop {} {\bibfield  {journal}
  {\bibinfo  {journal} {J. Phys. Chem B.}\ }\textbf {\bibinfo {volume} {108}},\
  \bibinfo {pages} {7286} (\bibinfo {year} {2004})}\BibitemShut {NoStop}%
\bibitem [{\citenamefont {Guerrero-Garc{\'i}a}\ \emph
  {et~al.}(2010)\citenamefont {Guerrero-Garc{\'i}a}, \citenamefont
  {Gonz{\'a}lez-Tovar}, \citenamefont {Ch{\'a}vez-P{\'a}ez},\ and\
  \citenamefont {Lozada-Cassou}}]{ivan10}%
  \BibitemOpen
  \bibfield  {author} {\bibinfo {author} {\bibfnamefont {G.~I.}\ \bibnamefont
  {Guerrero-Garc{\'i}a}}, \bibinfo {author} {\bibfnamefont {E.}~\bibnamefont
  {Gonz{\'a}lez-Tovar}}, \bibinfo {author} {\bibfnamefont {M.}~\bibnamefont
  {Ch{\'a}vez-P{\'a}ez}}, \ and\ \bibinfo {author} {\bibfnamefont
  {M.}~\bibnamefont {Lozada-Cassou}},\ }\href@noop {} {\bibfield  {journal}
  {\bibinfo  {journal} {J. Chem. Phys.}\ }\textbf {\bibinfo {volume} {132}},\
  \bibinfo {pages} {054903(1)} (\bibinfo {year} {2010})}\BibitemShut {NoStop}%
\bibitem [{\citenamefont {Chialvo}\ and\ \citenamefont
  {Simonson}(2008)}]{chialvo2008}%
  \BibitemOpen
  \bibfield  {author} {\bibinfo {author} {\bibfnamefont {A.~A.}\ \bibnamefont
  {Chialvo}}\ and\ \bibinfo {author} {\bibfnamefont {J.~M.}\ \bibnamefont
  {Simonson}},\ }\href@noop {} {\bibfield  {journal} {\bibinfo  {journal} {J.
  Phys. Chem. C}\ }\textbf {\bibinfo {volume} {112}},\ \bibinfo {pages} {19521}
  (\bibinfo {year} {2008})}\BibitemShut {NoStop}%
\bibitem [{\citenamefont {Lyklema}(2006)}]{lyklema06}%
  \BibitemOpen
  \bibfield  {author} {\bibinfo {author} {\bibfnamefont {J.}~\bibnamefont
  {Lyklema}},\ }\href@noop {} {\bibfield  {journal} {\bibinfo  {journal}
  {Colloids and Surfaces A: Physicochem. Eng. Aspects}\ }\textbf {\bibinfo
  {volume} {291}},\ \bibinfo {pages} {3} (\bibinfo {year} {2006})}\BibitemShut
  {NoStop}%
\bibitem [{\citenamefont {Messina}(2007)}]{Messina2007}%
  \BibitemOpen
  \bibfield  {author} {\bibinfo {author} {\bibfnamefont {R.}~\bibnamefont
  {Messina}},\ }\href@noop {} {\bibfield  {journal} {\bibinfo  {journal} {J.
  Chem. Phys. 127, 214901.}\ }\textbf {\bibinfo {volume} {127}},\ \bibinfo
  {pages} {214901} (\bibinfo {year} {2007})}\BibitemShut {NoStop}%
\bibitem [{\citenamefont {Wang}\ and\ \citenamefont {Ma}(2010)}]{wang01}%
  \BibitemOpen
  \bibfield  {author} {\bibinfo {author} {\bibfnamefont {Z.-Y.}\ \bibnamefont
  {Wang}}\ and\ \bibinfo {author} {\bibfnamefont {Y.-Q.}\ \bibnamefont {Ma}},\
  }\href@noop {} {\bibfield  {journal} {\bibinfo  {journal} {J. Chem. Phys.}\
  }\textbf {\bibinfo {volume} {133}},\ \bibinfo {pages} {064704(1)} (\bibinfo
  {year} {2010})}\BibitemShut {NoStop}%
\bibitem [{\citenamefont {Chialvo}\ and\ \citenamefont
  {Simonson}(2011)}]{chialvo2011}%
  \BibitemOpen
  \bibfield  {author} {\bibinfo {author} {\bibfnamefont {A.~A.}\ \bibnamefont
  {Chialvo}}\ and\ \bibinfo {author} {\bibfnamefont {J.~M.}\ \bibnamefont
  {Simonson}},\ }\href@noop {} {\bibfield  {journal} {\bibinfo  {journal}
  {Condensed Matter Physics}\ }\textbf {\bibinfo {volume} {14}},\ \bibinfo
  {pages} {33002(1)} (\bibinfo {year} {2011})}\BibitemShut {NoStop}%
\bibitem [{\citenamefont {Guerreo-Garc{\'i}a}\ and\ \citenamefont {de-la
  Cruz}(2014)}]{olvera2014}%
  \BibitemOpen
  \bibfield  {author} {\bibinfo {author} {\bibfnamefont {G.~I.}\ \bibnamefont
  {Guerreo-Garc{\'i}a}}\ and\ \bibinfo {author} {\bibfnamefont {M.~O.}\
  \bibnamefont {de-la Cruz}},\ }\href@noop {} {\bibfield  {journal} {\bibinfo
  {journal} {J. Phys. Chem. B}\ }\textbf {\bibinfo {volume} {118}},\ \bibinfo
  {pages} {8854} (\bibinfo {year} {2014})}\BibitemShut {NoStop}%
\bibitem [{\citenamefont {Naji}\ \emph {et~al.}(2014)\citenamefont {Naji},
  \citenamefont {Ghodrat}, \citenamefont {Komaie-Moghaddam},\ and\
  \citenamefont {Podgornik}}]{podgornik2014}%
  \BibitemOpen
  \bibfield  {author} {\bibinfo {author} {\bibfnamefont {A.}~\bibnamefont
  {Naji}}, \bibinfo {author} {\bibfnamefont {M.}~\bibnamefont {Ghodrat}},
  \bibinfo {author} {\bibfnamefont {H.}~\bibnamefont {Komaie-Moghaddam}}, \
  and\ \bibinfo {author} {\bibfnamefont {R.}~\bibnamefont {Podgornik}},\
  }\href@noop {} {\bibfield  {journal} {\bibinfo  {journal} {J. Chem. Phys.}\
  }\textbf {\bibinfo {volume} {141}},\ \bibinfo {pages} {174704} (\bibinfo
  {year} {2014})}\BibitemShut {NoStop}%
\bibitem [{\citenamefont {Wang}(2016)}]{wang2016}%
  \BibitemOpen
  \bibfield  {author} {\bibinfo {author} {\bibfnamefont {Z.-Y.}\ \bibnamefont
  {Wang}},\ }\href@noop {} {\bibfield  {journal} {\bibinfo  {journal} {J. Stat.
  Mech.}\ }\textbf {\bibinfo {volume} {2016}},\ \bibinfo {pages} {043205}
  (\bibinfo {year} {2016})}\BibitemShut {NoStop}%
\bibitem [{\citenamefont {Patra}(2016)}]{patra2016}%
  \BibitemOpen
  \bibfield  {author} {\bibinfo {author} {\bibfnamefont {C.~N.}\ \bibnamefont
  {Patra}},\ }\href@noop {} {\bibfield  {journal} {\bibinfo  {journal} {Mol.
  Phys.}\ }\textbf {\bibinfo {volume} {114}},\ \bibinfo {pages} {2341}
  (\bibinfo {year} {2016})}\BibitemShut {NoStop}%
\bibitem [{\citenamefont {Jang}\ \emph
  {et~al.}(2017{\natexlab{a}})\citenamefont {Jang}, \citenamefont {Shin},\ and\
  \citenamefont {Kim}}]{seanea2017b}%
  \BibitemOpen
  \bibfield  {author} {\bibinfo {author} {\bibfnamefont {S.}~\bibnamefont
  {Jang}}, \bibinfo {author} {\bibfnamefont {G.~R.}\ \bibnamefont {Shin}}, \
  and\ \bibinfo {author} {\bibfnamefont {S.-C.}\ \bibnamefont {Kim}},\ }\href
  {\doibase /10.1016/j.molliq.2017.03.121} {\bibfield  {journal} {\bibinfo
  {journal} {J. Mol. Liquids}\ }\textbf {\bibinfo {volume} {237}},\ \bibinfo
  {pages} {282} (\bibinfo {year} {2017}{\natexlab{a}})}\BibitemShut {NoStop}%
\bibitem [{\citenamefont {Bhuiyan}\ and\ \citenamefont
  {Outhwaite}(2017)}]{Bhuiyan2017}%
  \BibitemOpen
  \bibfield  {author} {\bibinfo {author} {\bibfnamefont {L.~B.}\ \bibnamefont
  {Bhuiyan}}\ and\ \bibinfo {author} {\bibfnamefont {C.~W.}\ \bibnamefont
  {Outhwaite}},\ }\href@noop {} {\bibfield  {journal} {\bibinfo  {journal}
  {Condensed Matter Physics}\ }\textbf {\bibinfo {volume} {20}},\ \bibinfo
  {pages} {33801} (\bibinfo {year} {2017})}\BibitemShut {NoStop}%
\bibitem [{\citenamefont {Kirkwood}\ and\ \citenamefont
  {Poirier}(1954)}]{poirier1954}%
  \BibitemOpen
  \bibfield  {author} {\bibinfo {author} {\bibfnamefont {J.~G.}\ \bibnamefont
  {Kirkwood}}\ and\ \bibinfo {author} {\bibfnamefont {J.~C.}\ \bibnamefont
  {Poirier}},\ }\href@noop {} {\bibfield  {journal} {\bibinfo  {journal} {J.
  Phys. Chem.}\ }\textbf {\bibinfo {volume} {591-596}},\ \bibinfo {pages}
  {1282} (\bibinfo {year} {1954})}\BibitemShut {NoStop}%
\bibitem [{\citenamefont {Deserno}\ \emph {et~al.}(2001)\citenamefont
  {Deserno}, \citenamefont {Jim{\'e}nez-{\'A}ngeles}, \citenamefont {Holm},\
  and\ \citenamefont {Lozada-Cassou}}]{deserno2001}%
  \BibitemOpen
  \bibfield  {author} {\bibinfo {author} {\bibfnamefont {M.}~\bibnamefont
  {Deserno}}, \bibinfo {author} {\bibfnamefont {F.}~\bibnamefont
  {Jim{\'e}nez-{\'A}ngeles}}, \bibinfo {author} {\bibfnamefont
  {C.}~\bibnamefont {Holm}}, \ and\ \bibinfo {author} {\bibfnamefont
  {M.}~\bibnamefont {Lozada-Cassou}},\ }\href@noop {} {\bibfield  {journal}
  {\bibinfo  {journal} {J. Phys. Chem. B}\ }\textbf {\bibinfo {volume} {105}},\
  \bibinfo {pages} {10983} (\bibinfo {year} {2001})}\BibitemShut {NoStop}%
\bibitem [{\citenamefont {Belloni}(1998)}]{belloni1998}%
  \BibitemOpen
  \bibfield  {author} {\bibinfo {author} {\bibfnamefont {L.}~\bibnamefont
  {Belloni}},\ }\href@noop {} {\bibfield  {journal} {\bibinfo  {journal}
  {Colloids Surfaces A: Physicochem. Eng. Aspects}\ }\textbf {\bibinfo {volume}
  {140}},\ \bibinfo {pages} {227} (\bibinfo {year} {1998})}\BibitemShut
  {NoStop}%
\bibitem [{\citenamefont {Bocquet}\ \emph {et~al.}(2002)\citenamefont
  {Bocquet}, \citenamefont {Trizac},\ and\ \citenamefont
  {Aubouy}}]{bocquet2002}%
  \BibitemOpen
  \bibfield  {author} {\bibinfo {author} {\bibfnamefont {L.}~\bibnamefont
  {Bocquet}}, \bibinfo {author} {\bibfnamefont {E.}~\bibnamefont {Trizac}}, \
  and\ \bibinfo {author} {\bibfnamefont {M.}~\bibnamefont {Aubouy}},\
  }\href@noop {} {\bibfield  {journal} {\bibinfo  {journal} {J. Chem. Phys.}\
  }\textbf {\bibinfo {volume} {117}},\ \bibinfo {pages} {8138} (\bibinfo {year}
  {2002})}\BibitemShut {NoStop}%
\bibitem [{\citenamefont {Ise}\ \emph {et~al.}(1999)\citenamefont {Ise},
  \citenamefont {Konishi},\ and\ \citenamefont {Tata}}]{ise99}%
  \BibitemOpen
  \bibfield  {author} {\bibinfo {author} {\bibfnamefont {N.}~\bibnamefont
  {Ise}}, \bibinfo {author} {\bibfnamefont {T.}~\bibnamefont {Konishi}}, \ and\
  \bibinfo {author} {\bibfnamefont {B.~V.~R.}\ \bibnamefont {Tata}},\
  }\href@noop {} {\bibfield  {journal} {\bibinfo  {journal} {Langmuir}\
  }\textbf {\bibinfo {volume} {15}},\ \bibinfo {pages} {4176} (\bibinfo {year}
  {1999})},\ \bibinfo {note} {and references therein}\BibitemShut {NoStop}%
\bibitem [{\citenamefont {Linse}\ and\ \citenamefont
  {Lobaskin}(1999)}]{linse99}%
  \BibitemOpen
  \bibfield  {author} {\bibinfo {author} {\bibfnamefont {P.}~\bibnamefont
  {Linse}}\ and\ \bibinfo {author} {\bibfnamefont {V.}~\bibnamefont
  {Lobaskin}},\ }\href@noop {} {\bibfield  {journal} {\bibinfo  {journal}
  {Phys. Rev. Lett.}\ }\textbf {\bibinfo {volume} {83}},\ \bibinfo {pages}
  {4208} (\bibinfo {year} {1999})}\BibitemShut {NoStop}%
\bibitem [{\citenamefont {Belloni}(1985)}]{belloni1985}%
  \BibitemOpen
  \bibfield  {author} {\bibinfo {author} {\bibfnamefont {L.}~\bibnamefont
  {Belloni}},\ }\href@noop {} {\bibfield  {journal} {\bibinfo  {journal}
  {Chemical Physics 99 (1985) 43-54}\ }\textbf {\bibinfo {volume} {99}},\
  \bibinfo {pages} {43} (\bibinfo {year} {1985})}\BibitemShut {NoStop}%
\bibitem [{\citenamefont {Lobaskin}\ and\ \citenamefont
  {Qamhieh}(2003)}]{lobaskin03}%
  \BibitemOpen
  \bibfield  {author} {\bibinfo {author} {\bibfnamefont {V.}~\bibnamefont
  {Lobaskin}}\ and\ \bibinfo {author} {\bibfnamefont {K.}~\bibnamefont
  {Qamhieh}},\ }\href@noop {} {\bibfield  {journal} {\bibinfo  {journal} {J.
  Phys. Chem. B}\ }\textbf {\bibinfo {volume} {107}},\ \bibinfo {pages} {8022}
  (\bibinfo {year} {2003})}\BibitemShut {NoStop}%
\bibitem [{\citenamefont {Gutiérrez-Valladares}\ \emph
  {et~al.}(2011)\citenamefont {Gutiérrez-Valladares}, \citenamefont
  {Luk\v{s}i\v{c}}, \citenamefont {Mill\'an-Malo}, \citenamefont {Hribar-Lee},\
  and\ \citenamefont {Vlachy}}]{vlachy2011}%
  \BibitemOpen
  \bibfield  {author} {\bibinfo {author} {\bibfnamefont {E.}~\bibnamefont
  {Gutiérrez-Valladares}}, \bibinfo {author} {\bibfnamefont {M.}~\bibnamefont
  {Luk\v{s}i\v{c}}}, \bibinfo {author} {\bibfnamefont {B.}~\bibnamefont
  {Mill\'an-Malo}}, \bibinfo {author} {\bibfnamefont {B.}~\bibnamefont
  {Hribar-Lee}}, \ and\ \bibinfo {author} {\bibfnamefont {V.}~\bibnamefont
  {Vlachy}},\ }\href@noop {} {\bibfield  {journal} {\bibinfo  {journal}
  {Condens. Matter Phys.}\ }\textbf {\bibinfo {volume} {14}},\ \bibinfo {pages}
  {33003} (\bibinfo {year} {2011})}\BibitemShut {NoStop}%
\bibitem [{\citenamefont {Manzanilla-Granados}\ \emph
  {et~al.}(2011{\natexlab{a}})\citenamefont {Manzanilla-Granados},
  \citenamefont {Jim{\'e}nez-{\'A}ngeles},\ and\ \citenamefont
  {Lozada-Cassou}}]{manzanilla2011a}%
  \BibitemOpen
  \bibfield  {author} {\bibinfo {author} {\bibfnamefont {H.~M.}\ \bibnamefont
  {Manzanilla-Granados}}, \bibinfo {author} {\bibfnamefont {F.}~\bibnamefont
  {Jim{\'e}nez-{\'A}ngeles}}, \ and\ \bibinfo {author} {\bibfnamefont
  {M.}~\bibnamefont {Lozada-Cassou}},\ }\href@noop {} {\bibfield  {journal}
  {\bibinfo  {journal} {Colloids and Surfaces A: Physicochem. Eng. Aspects}\
  }\textbf {\bibinfo {volume} {376}},\ \bibinfo {pages} {59} (\bibinfo {year}
  {2011}{\natexlab{a}})}\BibitemShut {NoStop}%
\bibitem [{\citenamefont {Manzanilla-Granados}\ \emph
  {et~al.}(2011{\natexlab{b}})\citenamefont {Manzanilla-Granados},
  \citenamefont {Jim{\'e}nez-{\'A}ngeles},\ and\ \citenamefont
  {Lozada-Cassou}}]{manzanilla2011b}%
  \BibitemOpen
  \bibfield  {author} {\bibinfo {author} {\bibfnamefont {H.~M.}\ \bibnamefont
  {Manzanilla-Granados}}, \bibinfo {author} {\bibfnamefont {F.}~\bibnamefont
  {Jim{\'e}nez-{\'A}ngeles}}, \ and\ \bibinfo {author} {\bibfnamefont
  {M.}~\bibnamefont {Lozada-Cassou}},\ }\href@noop {} {\bibfield  {journal}
  {\bibinfo  {journal} {J. Phys. Chem. B}\ }\textbf {\bibinfo {volume} {115}},\
  \bibinfo {pages} {12094} (\bibinfo {year} {2011}{\natexlab{b}})}\BibitemShut
  {NoStop}%
\bibitem [{\citenamefont {Manzanilla-Granados}\ and\ \citenamefont
  {Lozada-Cassou}(2013)}]{manzanilla2013}%
  \BibitemOpen
  \bibfield  {author} {\bibinfo {author} {\bibfnamefont {H.}~\bibnamefont
  {Manzanilla-Granados}}\ and\ \bibinfo {author} {\bibfnamefont
  {M.}~\bibnamefont {Lozada-Cassou}},\ }\href@noop {} {\bibfield  {journal}
  {\bibinfo  {journal} {J. Phys. Chem. B}\ }\textbf {\bibinfo {volume} {117}},\
  \bibinfo {pages} {11812} (\bibinfo {year} {2013})},\ \bibinfo {note} {see
  Fig. 9.}\BibitemShut {Stop}%
\bibitem [{\citenamefont {Tata}\ \emph {et~al.}(2008)\citenamefont {Tata},
  \citenamefont {Mohanty},\ and\ \citenamefont {Valsakumara}}]{tata2008}%
  \BibitemOpen
  \bibfield  {author} {\bibinfo {author} {\bibfnamefont {B.}~\bibnamefont
  {Tata}}, \bibinfo {author} {\bibfnamefont {P.}~\bibnamefont {Mohanty}}, \
  and\ \bibinfo {author} {\bibfnamefont {M.}~\bibnamefont {Valsakumara}},\
  }\href@noop {} {\bibfield  {journal} {\bibinfo  {journal} {Solid State
  Communications}\ }\textbf {\bibinfo {volume} {147}},\ \bibinfo {pages} {360}
  (\bibinfo {year} {2008})},\ \bibinfo {note} {and references
  therein}\BibitemShut {NoStop}%
\bibitem [{\citenamefont {M{\"u}nster}(1974)}]{munster1}%
  \BibitemOpen
  \bibfield  {author} {\bibinfo {author} {\bibfnamefont {A.}~\bibnamefont
  {M{\"u}nster}},\ }\href@noop {} {\emph {\bibinfo {title} {Statistical
  Thermodynamics}}},\ \bibinfo {edition} {first english}\ ed.,\ Vol.\ \bibinfo
  {volume} {I and II}\ (\bibinfo  {publisher} {Springer-Verlag},\ \bibinfo
  {address} {Berlin},\ \bibinfo {year} {1974})\BibitemShut {NoStop}%
\bibitem [{\citenamefont {McQuarrie}(1976)}]{McQuarrie}%
  \BibitemOpen
  \bibfield  {author} {\bibinfo {author} {\bibfnamefont {D.~A.}\ \bibnamefont
  {McQuarrie}},\ }\href@noop {} {\emph {\bibinfo {title} {Statistical
  Mechanics}}}\ (\bibinfo  {publisher} {Harper and Row},\ \bibinfo {address}
  {New York},\ \bibinfo {year} {1976})\BibitemShut {NoStop}%
\bibitem [{\citenamefont {Lozada-Cassou}(1992)}]{Lozada-Cassou92a}%
  \BibitemOpen
  \bibfield  {author} {\bibinfo {author} {\bibfnamefont {M.}~\bibnamefont
  {Lozada-Cassou}},\ }in\ \href@noop {} {\emph {\bibinfo {booktitle}
  {Fundamentals of Inhomogeneous Fluds}}},\ \bibinfo {editor} {edited by\
  \bibinfo {editor} {\bibfnamefont {D.}~\bibnamefont {Henderson}}}\ (\bibinfo
  {publisher} {Marcel Dekker},\ \bibinfo {address} {New York},\ \bibinfo {year}
  {1992})\ Chap.~\bibinfo {chapter} {8}\BibitemShut {NoStop}%
\bibitem [{\citenamefont {Vlachy}\ and\ \citenamefont
  {Haymet}(1986)}]{vlachy86}%
  \BibitemOpen
  \bibfield  {author} {\bibinfo {author} {\bibfnamefont {V.}~\bibnamefont
  {Vlachy}}\ and\ \bibinfo {author} {\bibfnamefont {A.~D.~J.}\ \bibnamefont
  {Haymet}},\ }\href@noop {} {\bibfield  {journal} {\bibinfo  {journal} {J.
  Chem Phys.}\ }\textbf {\bibinfo {volume} {84}},\ \bibinfo {pages} {5874}
  (\bibinfo {year} {1986})}\BibitemShut {NoStop}%
\bibitem [{\citenamefont {Hribar}\ \emph {et~al.}(2000)\citenamefont {Hribar},
  \citenamefont {Vlachy}, \citenamefont {Bhuiyan},\ and\ \citenamefont
  {Outhwaite}}]{vlachy2000}%
  \BibitemOpen
  \bibfield  {author} {\bibinfo {author} {\bibfnamefont {B.}~\bibnamefont
  {Hribar}}, \bibinfo {author} {\bibfnamefont {V.}~\bibnamefont {Vlachy}},
  \bibinfo {author} {\bibfnamefont {L.~B.}\ \bibnamefont {Bhuiyan}}, \ and\
  \bibinfo {author} {\bibfnamefont {C.~W.}\ \bibnamefont {Outhwaite}},\
  }\href@noop {} {\bibfield  {journal} {\bibinfo  {journal} {J. Phys. Chem. B}\
  }\textbf {\bibinfo {volume} {104}},\ \bibinfo {pages} {11522} (\bibinfo
  {year} {2000})}\BibitemShut {NoStop}%
\bibitem [{\citenamefont {Yu}\ \emph {et~al.}(2004)\citenamefont {Yu},
  \citenamefont {Wu},\ and\ \citenamefont {Gao}}]{Wu2004}%
  \BibitemOpen
  \bibfield  {author} {\bibinfo {author} {\bibfnamefont {Y.-X.}\ \bibnamefont
  {Yu}}, \bibinfo {author} {\bibfnamefont {J.}~\bibnamefont {Wu}}, \ and\
  \bibinfo {author} {\bibfnamefont {G.-H.}\ \bibnamefont {Gao}},\ }\href@noop
  {} {\bibfield  {journal} {\bibinfo  {journal} {J. Chem. Phys.}\ }\textbf
  {\bibinfo {volume} {120}},\ \bibinfo {pages} {7223} (\bibinfo {year}
  {2004})}\BibitemShut {NoStop}%
\bibitem [{\citenamefont {Quesada-P{\'e}rez}\ and\ \citenamefont
  {Hidalgo-{\'A}lvarez}(2006)}]{molina2006}%
  \BibitemOpen
  \bibfield  {author} {\bibinfo {author} {\bibfnamefont {A.~M. M.~M.}\
  \bibnamefont {Quesada-P{\'e}rez}}\ and\ \bibinfo {author} {\bibfnamefont
  {R.}~\bibnamefont {Hidalgo-{\'A}lvarez}},\ }\href@noop {} {\bibfield
  {journal} {\bibinfo  {journal} {J. Phys. Chem. B}\ }\textbf {\bibinfo
  {volume} {110}},\ \bibinfo {pages} {1326} (\bibinfo {year}
  {2006})}\BibitemShut {NoStop}%
\bibitem [{\citenamefont {Guerreo-Garc{\'i}a}\ \emph
  {et~al.}(2011)\citenamefont {Guerreo-Garc{\'i}a}, \citenamefont
  {Gonz{\'a}les-Tovar},\ and\ \citenamefont {de-la Cruz}}]{olvera2011}%
  \BibitemOpen
  \bibfield  {author} {\bibinfo {author} {\bibfnamefont {G.~I.}\ \bibnamefont
  {Guerreo-Garc{\'i}a}}, \bibinfo {author} {\bibfnamefont {E.}~\bibnamefont
  {Gonz{\'a}les-Tovar}}, \ and\ \bibinfo {author} {\bibfnamefont {M.~O.}\
  \bibnamefont {de-la Cruz}},\ }\href@noop {} {\bibfield  {journal} {\bibinfo
  {journal} {J. Chem. Phys.}\ }\textbf {\bibinfo {volume} {135}},\ \bibinfo
  {pages} {054701} (\bibinfo {year} {2011})}\BibitemShut {NoStop}%
\bibitem [{\citenamefont {Jang}\ \emph
  {et~al.}(2017{\natexlab{b}})\citenamefont {Jang}, \citenamefont {Shin},\ and\
  \citenamefont {Kim}}]{seanea2017}%
  \BibitemOpen
  \bibfield  {author} {\bibinfo {author} {\bibfnamefont {S.}~\bibnamefont
  {Jang}}, \bibinfo {author} {\bibfnamefont {G.~R.}\ \bibnamefont {Shin}}, \
  and\ \bibinfo {author} {\bibfnamefont {S.-C.}\ \bibnamefont {Kim}},\ }\href
  {\doibase /10.1080/00268976.2017.1321156} {\bibfield  {journal} {\bibinfo
  {journal} {Mol. Phys.}\ }\textbf {\bibinfo {volume} {115}},\ \bibinfo {pages}
  {2411} (\bibinfo {year} {2017}{\natexlab{b}})}\BibitemShut {NoStop}%
\bibitem [{\citenamefont {Friedman}(1985)}]{Friedmanbook}%
  \BibitemOpen
  \bibfield  {author} {\bibinfo {author} {\bibfnamefont {H.~L.}\ \bibnamefont
  {Friedman}},\ }\href@noop {} {\emph {\bibinfo {title} {A course in
  Statistical Mechanics}}}\ (\bibinfo  {publisher} {Prentice-Hall, Inc.},\
  \bibinfo {year} {1985})\BibitemShut {NoStop}%
\bibitem [{\citenamefont {Hiroike}(1977)}]{kazuo77}%
  \BibitemOpen
  \bibfield  {author} {\bibinfo {author} {\bibfnamefont {K.}~\bibnamefont
  {Hiroike}},\ }\href@noop {} {\bibfield  {journal} {\bibinfo  {journal} {Mol.
  Phys.}\ }\textbf {\bibinfo {volume} {33}},\ \bibinfo {pages} {1195} (\bibinfo
  {year} {1977})}\BibitemShut {NoStop}%
\bibitem [{\citenamefont {Lobaskin}\ and\ \citenamefont
  {Linse}(1999)}]{lobaskin99}%
  \BibitemOpen
  \bibfield  {author} {\bibinfo {author} {\bibfnamefont {V.}~\bibnamefont
  {Lobaskin}}\ and\ \bibinfo {author} {\bibfnamefont {P.}~\bibnamefont
  {Linse}},\ }\href@noop {} {\bibfield  {journal} {\bibinfo  {journal} {J.
  Chem. Phys.}\ }\textbf {\bibinfo {volume} {111}},\ \bibinfo {pages} {4300}
  (\bibinfo {year} {1999})}\BibitemShut {NoStop}%
\bibitem [{\citenamefont {Ise}\ \emph {et~al.}(1983)\citenamefont {Ise},
  \citenamefont {Okubo}, \citenamefont {Sugimura}, \citenamefont {Ito},\ and\
  \citenamefont {Nolte}}]{ise83}%
  \BibitemOpen
  \bibfield  {author} {\bibinfo {author} {\bibfnamefont {N.}~\bibnamefont
  {Ise}}, \bibinfo {author} {\bibfnamefont {T.}~\bibnamefont {Okubo}}, \bibinfo
  {author} {\bibfnamefont {M.}~\bibnamefont {Sugimura}}, \bibinfo {author}
  {\bibfnamefont {K.}~\bibnamefont {Ito}}, \ and\ \bibinfo {author}
  {\bibfnamefont {H.~J.}\ \bibnamefont {Nolte}},\ }\href@noop {} {\bibfield
  {journal} {\bibinfo  {journal} {J. chem. Phys}\ }\textbf {\bibinfo {volume}
  {78}},\ \bibinfo {pages} {536} (\bibinfo {year} {1983})}\BibitemShut
  {NoStop}%
\end{thebibliography}

%

\end{document}